\documentclass[%
 reprint,
 amsmath,amssymb,
 aps,
]{revtex4-2}

\usepackage[utf8]{inputenc}

\usepackage{amssymb}

\usepackage{amsmath}

\usepackage{amsfonts}

\usepackage{pifont}

\usepackage[usenames,dvipsnames]{xcolor}

\usepackage[pdftex]{graphicx}

\usepackage{bm}

\usepackage{hyperref}

\hypersetup{plainpages=false,colorlinks=true,linkcolor=blue, citecolor=blue, urlcolor=blue}

\usepackage{array}

\newcolumntype{L}[1]{>{\raggedright\let\newline\\\arraybackslash\hspace{0pt}}m{#1}}

\newcolumntype{C}[1]{>{\centering\let\newline\\\arraybackslash\hspace{0pt}}m{#1}}

\newcolumntype{R}[1]{>{\raggedleft\let\newline\\\arraybackslash\hspace{0pt}}m{#1}}

\newcommand{\dd}{\text{d}}

\usepackage{pifont}

\newcommand{\hham}{\mathcal{H}}
\newcommand{\Gmatsu}{\mathcal{G}}

\newcommand{\Kbf}{\mathbf{K}}%
\newcommand{\kbf}{\mathbf{k}}%
\newcommand{\qbf}{\mathbf{q}}%

\begin{document}

\title{The Frenkel line and the pseudogap: an analogy between classical and electronic fluids 
}

\author{J. \surname{Fournier}$^{1}$}

\author{P.-O. \surname{Downey}$^1$}

\author{O. \surname{Gingras}$^{2,3}$}

\author{C.-D.  \surname{Hébert}$^1$}

\author{M. \surname{Charlebois}$^{1,4}$}

\author{A.-M. S. \surname{Tremblay}$^1$}

\affiliation{$^1$D\'epartement de physique, RQMP and Institut quantique, Universit\'e de Sherbrooke, Qu\'ebec, Canada J1K 2R1}
\affiliation{$^2$Center for Computational Quantum Physics, Flatiron Institute, 162 Fifth Avenue, New York, New York 10010, USA}
\affiliation{$^3$Université Paris-Saclay, CNRS, CEA, Institut de physique théorique, 91191, Gif-sur-Yvette, France}
\affiliation{$^4$D\'epartement de Biochimie, Chimie, Physique et Science Forensique, Institut de Recherche sur l’Hydrog\`ene, Universit\'e du Qu\'ebec \`a Trois-Rivi\`eres, Trois-Rivi\`eres, Qu\'ebec G9A 5H7, Canada}

\date{\today}
\begin{abstract}
    Asymptotically close to critical end-points of first-order transitions, maxima in thermodynamic quantities occur along a line called the Widom line, a concept first introduced in classical fluids.  
    This concept has been extended to strongly correlated electronic fluids in the context of the Mott transition.
    Namely, upon increasing interaction strength in the Hubbard model at half-filling, one finds a first-order Mott metal-insulator transition with a critical endpoint at high temperature, above which several crossover lines are observable. 
    Using the dynamical cluster approximation for the triangular-lattice Hubbard model, we compute a new crossover line, the Frenkel line, a concept borrowed from classical fluids that is useful for defining a sharp crossover between the pseudogap and the correlated Fermi liquid.
    The Frenkel line in the electron fluid is defined by the appearance of back-scattering upon entering the pseudogap. 
    The signature of back-scattering is the existence of a negative value in the time-domain optical conductivity.
    The Frenkel line extends to high temperatures.
\end{abstract}

\maketitle

\section{Introduction}
Thermodynamics and statistical physics provide conceptual frameworks that can shed light on phenomena that appear different. 
%
%
In the supercritical regime, the region found above the critical end-point of first-order transitions, crossover phenomena are observed \cite{georges_hubbard_1992,brazhkin_widom_2011,brazhkin_liquid-gas_2013,bolmatov_frenkel_2015}.
The line towards which maxima in thermodynamic response functions converge as the critical end-point is approached above the first-order Mott metal-insulator transition of the electronic fluid was shown to be analogous to the so-called Widom line that appears in the very different context of the liquid-gas transition~\cite{sordi_pseudogap_2012,downey_mott_2023, downey_doping_2024,brazhkin_widom_2011,simeoni_widom_2010}. This establishes an analogy between the electronic fluid and classical fluids.
This analogy also holds in the phase diagram of quantum chromodynamics~\cite{Sordi_Tremblay_2024}.
Here we focus on the crossover between the pseudogap and the metallic phase in the supercritical region above the first-order Mott transition, a phase transition that plays a key role in the physics of strongly correlated materials \cite{georges_hubbard_1992,Jarrell_1992, lefebvre2000mott, Morosan_Natelson_Nevidomskyy_Si_2012, dang_mott_2015,Pustogow_Bories_Lohle_Rosslhuber_Zhukova_Gorshunov_Tomic_Schlueter_Hubner_Hiramatsu:2018,downey_mott_2023,chatzieleftheriou_mott_2023,wietek_mott_2021, demedici_orbital-selective_2005,eisenlohr_mott_2019,senthil_theory_2008,Fratino-Rozenberg:2021,Pustogow_Saito_Löhle_Sanz_Alonso_Kawamoto_Dobrosavljevic_Dressel_Fratini_2021}. 

Currently, several different criteria are used to define the border between the pseudogap and the metal. Experimentally, one approach is to examine the angle-resolved photoemission spectroscopy and track the temperature at which the quasiparticle peak weakens or disappears at certain locations on the Fermi surface, signalling a suppression of low-energy excitations~\cite{loeser1996excitation,armitage2001anomalous,norman_pseudogap_2005,tremblay_pseudogap_2006,senechal_hot_2004,timusk_pseudogap_1999,fernandes2012preemptive,Kaminski_Kondo_Takeuchi_Gu_2015,downey_mott_2023,imajo2023pseudogap,menke_superconductivity_2024}. Another approach is to determine when the Drude peak centered at zero frequency disappears in the optical conductivity. This suggests a deviation from Drude’s theory, indicating that quasiparticles are no longer well-defined~\cite{orenstein_frequency-_1990,homes_optical_1993,timusk_pseudogap_1999,Pustogow_Bories_Lohle_Rosslhuber_Zhukova_Gorshunov_Tomic_Schlueter_Hubner_Hiramatsu:2018,PhysRevB.44.6909,pound2011optical,PhysRevB.73.174501,PhysRevB.86.100501,lobo_optical_2011,PhysRevB.69.024504,PhysRevLett.87.217001,PhysRevB.97.045117,drichko2007drude,hwang2008manifestation}. 
As a final example, in nuclear magnetic resonance, the fall of the Knight shift with decreasing temperature was the first indication of the pseudogap in cuprates~\cite{Alloul_NMR:1988,frachet2020hidden}.


In classical fluids, a dynamic crossover line dubbed the Frenkel line has recently been proposed to separate the rigid (liquid-like) and nonrigid (gas-like) regimes in the supercritical state \cite{brazhkin_two_2012,brazhkin_liquid-gas_2013,bolmatov_frenkel_2015,bolmatov_unified_2015,trachenko_theory_2023,pruteanu_frenkel_2021,yang_frenkel_2015}.
Just as the Widom line has been a useful concept in electronic fluids, here we show that an analog of the Frenkel line exists and is a useful concept in electronic fluids.
%

But why look for another crossover line? 
First, a key limitation of the above crossover lines is that they focus mainly on low-frequency behavior, potentially overlooking other spectral features of the pseudogap at higher energies.
Second, we suggest, as in Ref.~\cite{Sordi_Tremblay_2024}, that when the critical endpoint is hidden below a broken-symmetry phase or is otherwise inaccessible, measuring several crossover lines associated with different observables is useful.
Indeed, since the lines of maxima of various thermodynamic quantities defining most crossover lines should converge to a single line, the Widom line~\cite{brazhkin_widom_2011, sordi_pseudogap_2012}, linear scaling theory~\cite{Luo_PRL_Widom_scaling_2014} will help to find the location of the critical endpoint as well as the slope of the first-order transition.
%
Third, not all crossover lines extend to very high temperatures. 
To extrapolate the existence of a critical endpoint from high-temperature measurements, it is advantageous that some crossovers persist to high temperatures.
This property was shown for the Frenkel line in the case of fluids~\cite{brazhkin_widom_2011}, and it applies in our case, as well. 


The Frenkel line in classical fluids is defined with time-dependent dynamic quantities~\cite{bolmatov_frenkel_2015, pruteanu_frenkel_2021,trachenko_theory_2023} such as the onset of oscillations in the time-dependent velocity-velocity autocorrelation function~\cite{brazhkin_two_2012}.
In the electronic fluid, we claim that an analog quantity defines a crossover line between the Fermi liquid and the pseudogap in the supercritical regime of the Mott transition.  
The analogy with the classical fluids suggests studying the time-dependent current-current autocorrelation function. 

However, instead of directly computing the current-current correlation function $\chi_{jj}$, we focus on the real part of the optical conductivity, which is related to it through $\sigma_1(\omega) = \chi_{jj}''(\omega, \mathbf{q = 0})/\omega$ and is more readily accessible experimentally.
Here, we investigate the crossover between pseudogap and correlated Fermi-liquid instead of the rigid non-rigid crossover; a new criterion is thus needed to define the Frenkel line. 
We suggest that the Frenkel line is crossed when oscillations in the time-dependent Fourier transform $\sigma_1(t)$ of the frequency-dependent conductivity $\sigma_1(\omega)$ are sufficiently strong to lead to negative $\sigma_1(t)$. 

Let us explain this. The time-dependent retarded conductivity $\sigma^R(t)$ is obtained simply from the Fourier transform of the frequency-dependent one (Eq.~(A1) of Appendix~A). 
It is related to the response of the electrical current to a time-dependent electric field through $j(t)=\int \sigma^R(t-t')E(t')dt'$.
Since one can prove~(Appendix~\ref{sec:FT}) that $\sigma^R(t)=2\sigma_1(t)\Theta(t)$, where $\sigma_1(t)$ is the Fourier transform of the experimentally measured frequency-dependent real part of the optical conductivity, $\sigma_1(\omega)$, from now on we focus on $\sigma_1(t)$.

In a simple metal, the optical conductivity is dominated by a Drude peak, yielding a strictly positive and monotonically decaying $\sigma_1(t)$.
In the pseudogap regime, the transfer of spectral weight away from the Fermi level redistributes part of the optical conductivity from the low-frequency regime to the mid-infrared.
This can lead to negative $\sigma_1(t-t')$ for some interval of $t-t'$, indicating that a mid-infrared peak can cause the current at time $t$ to have contributions from the electric field at former times $t'$ that are in the opposite direction to the current at time $t$. 
This can be understood as a consequence of back-scattering. 
Therefore, the appearance of $\sigma_1(t)<0$ is an indication that the metallic Drude-like transport is disappearing, which in turn implies that the system is no longer in a correlated Fermi-liquid state. 
The redistribution of the spectral weight away from the Fermi level is responsible for the decrease of the DOS at the Fermi level and for the negative $\sigma_1(t)$. Both are consequences of the same dynamical transition.

\begin{figure}
    \centering
    \includegraphics[width=\linewidth]{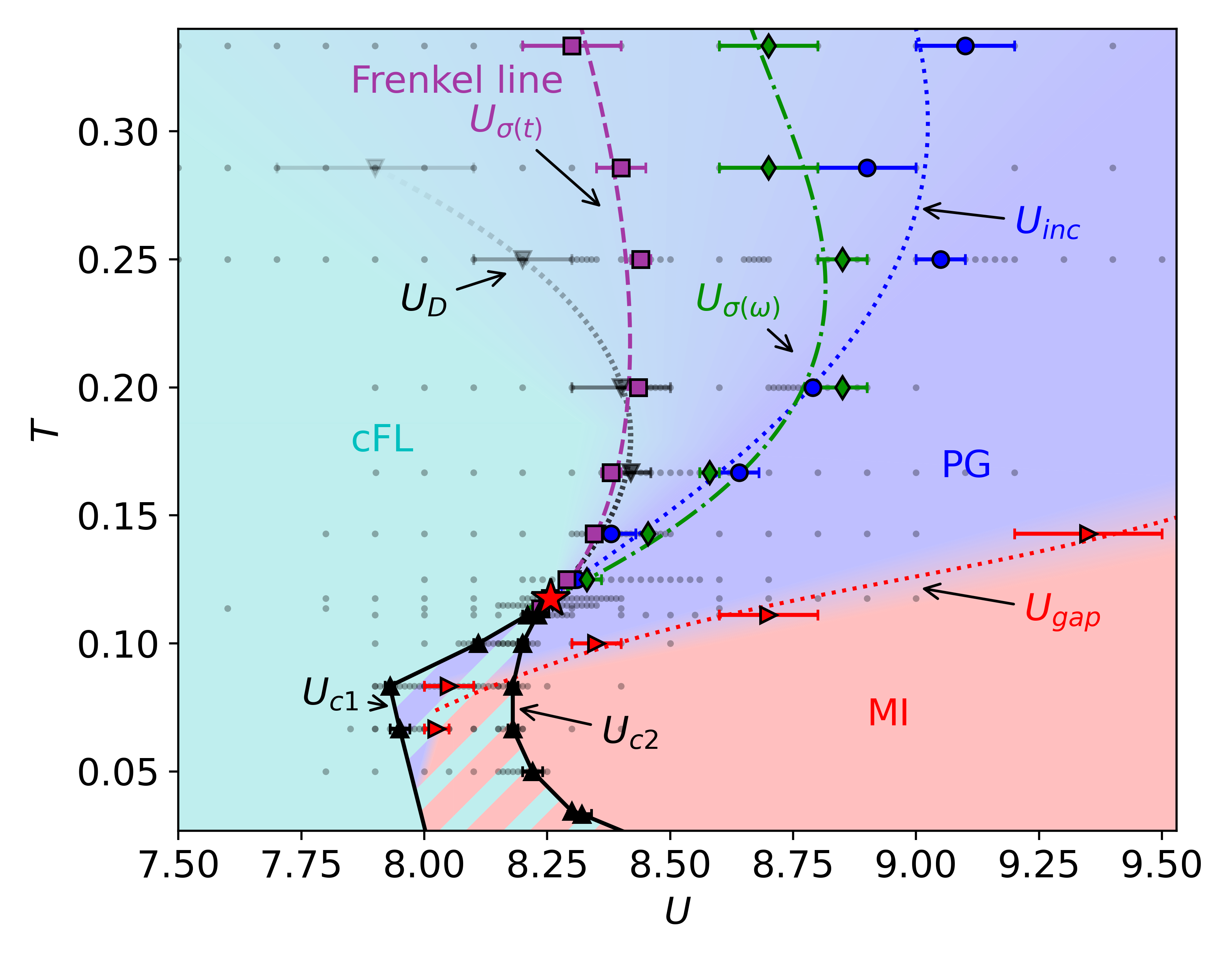}
    \caption{$T-U$ phase diagram of the isotropic triangular-lattice Hubbard model at half-filling. 
    The Frenkel line and the dynamic line $U_{\sigma(\omega)}$ have been added to the phase diagram reported in Ref.~\citenum{downey_mott_2023}, which already included the other crossover and transition lines.
    The dashed and the dotted lines represent crossover lines in the supercritical regime. The inflection in double occupancy $U_D$, the Frenkel line $U_{\sigma(t)}$, and the dynamic lines $U_{\sigma(\omega)}$ and $U_{inc}$ are respectively black, purple, green, and blue. 
    The markers for the different crossover lines between the pseudogap and correlated Fermi liquid are, respectively, inverted triangles, squares, diamonds, and circles. 
    The crossover lines $U_D$ ($U_W$ in the earlier notation) and $U_{inc}$ have been introduced in Ref.~\citenum{downey_mott_2023} along with the red dotted line $U_\text{gap}$ defined by the formation of a gap at $\omega=0$. 
    The correlated Fermi liquid (cFL), the pseudogap, and the Mott insulator regime are colored cyan, blue, and red.
    The red star at $U = 8.258$ and $T = 2/17$ identifies the Mott critical point. 
    The solid lines represent the first-order phase transition. 
    The hatched region between $U_{c1}$ and $U_{c2}$ represents a region of phase coexistence between the correlated Fermi liquid and the pseudogap or Mott insulating phase.  
    The tiny grey circles are the data points.
    }
    \label{fig:phasediag}
\end{figure}

With these concepts in mind, we now turn to simulations to provide a theoretical example of this general phenomenon.
We present results for the metal-pseudogap crossover above the Mott transition in the isotropic triangular lattice Hubbard model at half-filling.

The choice of the triangular lattice is motivated by the need to avoid finite-temperature long-wavelength fluctuations that would influence the finite-temperature critical endpoint of the first-order Mott transition~\cite{reymbaut2020motttransitionhightemperaturecrossovers}, which is governed by local physics. 
Layered organic materials~\cite{pustogow_thirty-year_2022} on a triangular lattice show the Mott transition and critical endpoint at finite temperature.

There have been many theoretical proposals for the ground state of the triangular lattice Hubbard model.
Because of the geometric frustration intrinsic to the triangular lattice, {\it for values of the interaction $U$ close to the critical value for the Mott transition}, all of these theories are consistent with a non-magnetic insulating ground state~\cite{wietek_mott_2021,morita_nonmagnetic_2002, kyung_mott_2006, sahebsara_hubbard_2008, laubachPhaseDiagramHubbard2015a, Misumi_Mott_triangular:2017, Yoshioka_triangular:2009, yang_effective_2010, szasz_chiral_2020, Tocchio_Montorsi_Becca_2021, PhysRevB.106.094420}. 
Recent calculations find that chiral spin-liquid or nematic phases can occur at finite temperature if time-reversal symmetry is broken~\cite{szasz_chiral_2020, Tocchio_Montorsi_Becca_2021, PhysRevB.106.094420}. 
They estimate that an upper bound is $T_c\sim0.0125$ in units of hopping~\cite{wietek_mott_2021}, consistent with the very small value of the $T=0$ gap~\cite{szasz_chiral_2020}, $\Delta\simeq 0.0065$. 
Fig.~6 of Ref.~\cite{wietek_mott_2021} shows that fluctuations associated with the chiral, nematic, stripy antiferromagnetic and spiral spin orders are small for values of $U$ near the Mott transition and temperatures larger than the Mott critical temperature~$T_c\sim 1/8.5 = 0.118$.
%
%

Our results, presented in the phase diagram in Fig.~\ref{fig:phasediag}, adapted from Ref.~\citenum{downey_mott_2023} along with several crossover lines identified in that work, demonstrate:
\renewcommand{\theenumi}{\roman{enumi}}
\begin{enumerate}
    \item the existence of an electronic Frenkel line defining the pseudogap-correlated Fermi liquid crossover exhibited by a strongly correlated electronic system,
    \item that this line, for all temperatures, is found in proximity to other lines that usually define that crossover,
    \item that the electronic Frenkel line, along with the other crossover lines, all converge towards the critical endpoint. 
\end{enumerate}
%
Since the Widom line is usually close to the loci of maximum isothermal compressibility $\kappa_{max}$~\cite{Luo_PRL_Widom_scaling_2014},
that quantity is often identified with the Widom line~\cite{sordi_pseudogap_2012,fratino_organizing_2016}. 
At half-filling, it is the line of inflection points of the double occupancy as a function of interaction that has played that role~\cite{downey_mott_2023}. 
Here, we take the more usual definition of the Widom line~\cite{Xu_Kumar_Buldyrev_Chen_Poole_Sciortino_Stanley_2005} as the line of convergence of supercritical crossover lines.
We do not draw that line, an asymptotic concept. 

\section{Model and method}
\begin{figure}
    \centering
    \includegraphics[width=0.98\linewidth]{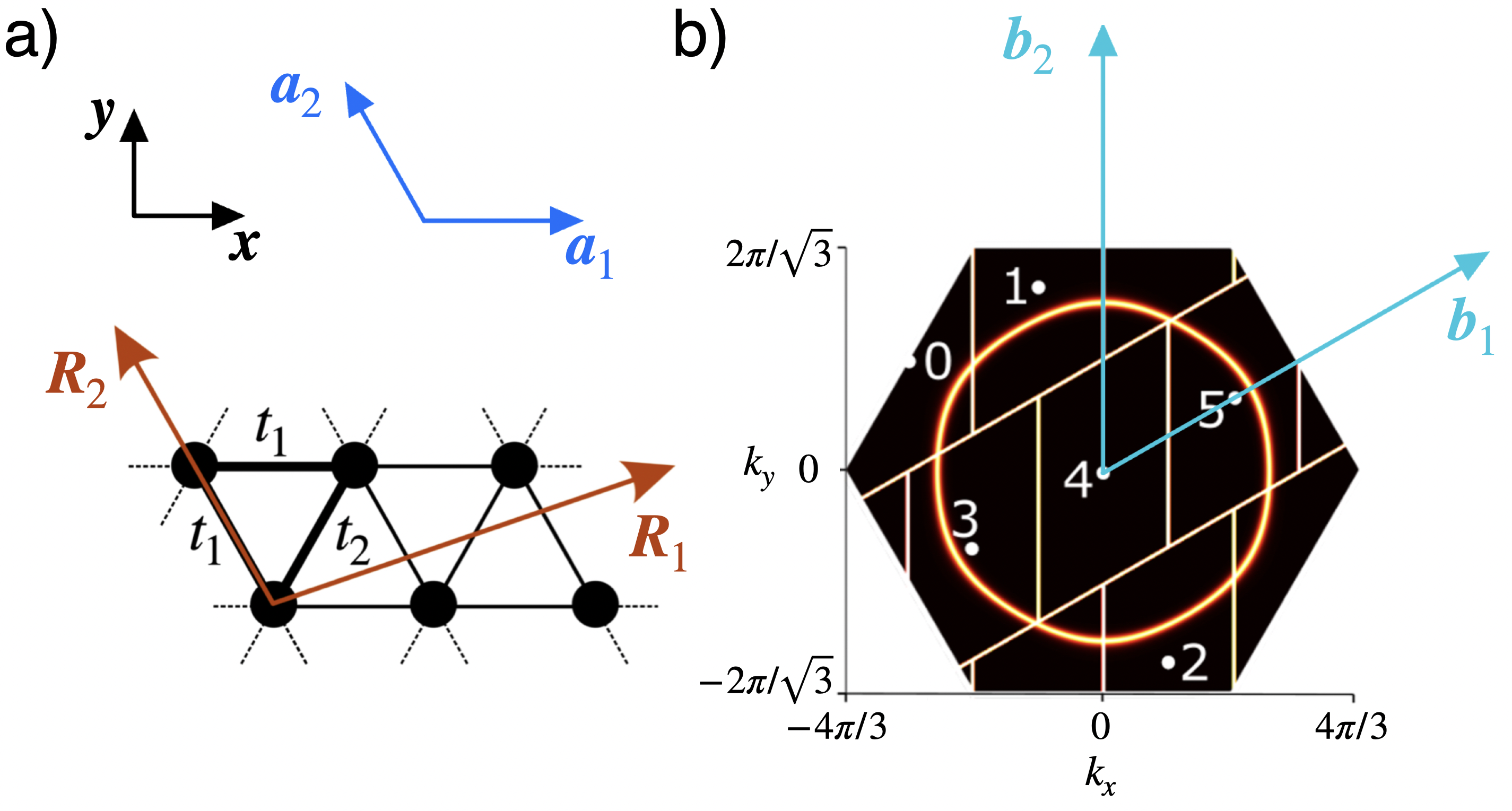}
    \caption{a) Six-site cluster with hopping terms 
    The coordinate axes $\mathbf{x}$ and $\mathbf{y}$ are shown in black.
    The primitive vectors $\mathbf{a}_1$ and $\mathbf{a}_2$ of the lattice are indicated in blue.
    The superlattice vectors $\mathbf{R}_1$ and $\mathbf{R}_2$ illustrate the periodic boundary conditions. 
    They are chosen such that if $t_2$ were zero, the resulting lattice would be bipartite.
    b)~Brillouin zone. The Fermi surface in orange is for $U = 0$ and $n = 1$ at $T = 0.1$ on the isotropic triangular lattice. 
    It is a hole
    Fermi surface. 
    The primitive reciprocal-lattice vectors $\mathbf{b}_1$ and $\mathbf{b}_2$ are shown in cyan.
    The patches for the DCA are outlined with their associated labels.  
    The reciprocal lattice vectors of the superlattice determine the choice of the patches in the Brillouin zone.
    The tilt of the patches arises because of a change of basis. In the computational reciprocal-lattice basis that we use, the patches are rectangular.
    %
    %
    }
    \label{fig:clusterBZ}
\end{figure}

The Hubbard model on the triangular lattice \cite{hubbard_electron_1963,sahebsara_hubbard_2008,qin_hubbard_2022,schafer_tracking_2021} is given by:
\begin{align}
    \hham =  -\sum_{  i,j, \sigma} t_{ij}c_{i\sigma}^\dagger c_{j\sigma}  + U \sum_i n_{i\uparrow}n_{i\downarrow} - \mu \sum_{i\sigma} n_{i\sigma}, \label{H_hubbard}
\end{align}
where creation-annihilation operators and occupation numbers are defined as usual, 
$U$ is the repulsive onsite interaction, $\mu$ is the chemical potential and $t_{ij}$ are hopping parameters between nearest-neighbor lattice sites as illustrated in Fig.~\ref{fig:clusterBZ}. 
We work at half-filling on the isotropic triangular lattice, setting $t_1$ and $t_2$ illustrated in the figure to $t_1=t_2=-1$.
We use two symbols on the isotropic triangular lattice, even if they are equal, for the sake of generality. 
The sign of $t_1$ in the dispersion relation 
\begin{multline}
    \epsilon_\kbf = -2t_1 \left[\cos(k_x)+ \cos\left(-\frac{k_x}{2}+ \frac{\sqrt{3}k_y}{2} \right)\right]\\-2t_2\cos\left(\frac{k_x}{2}+ \frac{\sqrt{3}k_y}{2} \right).
\end{multline}%
can be flipped by the shift of coordinates in the Brillouin zone $k_x\rightarrow k_x+\pi$, $k_y\rightarrow k_y - \frac{1}{\sqrt{3}}\pi$. 
More importantly, a spin-independent particle-hole transformation preserves the half-filling condition and changes the sign of both $t_1$ and $t_2$. 
This is how one can connect our results to those with $|t_1|=|t_2|$ and various signs of $t_1$ and $t_2$. 
Our crossover lines are independent of the sign of $t_1$ and $t_2$. 
%
%
%
We use the natural units $k_B=1$, $\hbar=1$, and lattice spacing $a=1$. 
The energy unit is the nearest-neighbor hopping parameter $|t_1|=1$.

We solve this model using the dynamical cluster approximation (DCA)~\cite{hettler2000dynamical,maier_quantum_2005,kotliar2006electronic,tremblay_pseudogap_2006}, which is a cluster extension of dynamical mean-field theory (DMFT)~\cite{georges_hubbard_1992, Jarrell_1992}. 
A cluster is necessary to observe a momentum-dependent pseudogap.

The auxiliary-field continuous-time quantum Monte-Carlo (CT-AUX) is the impurity solver for our six-site cluster.
\cite{gull_continuous-time_2011,werner_continuous-time_2006,gull_continuous-time_2008,semon_continuous-time_2014,semon2014lazy,chatzieleftheriou_mott_2023}.
In DCA, the Brillouin zone is segmented into $N_c$ patches, where $N_c$ is the number of sites in the cluster. 
Within these patches, the self-energy is constant. 
This makes DCA a coarse-grained method in $\textbf{k}$-space. 
Figure~\ref{fig:clusterBZ}a) shows our cluster in position space, and Fig.~\ref{fig:clusterBZ}b) the corresponding six patches in momentum space. 
%

One might argue that we need an exact solution of the many-body problem to prove our point.
The DCA becomes exact in the limit where $N_c \rightarrow\infty$.
Ref.~\cite{downey_mott_2023} studied the Mott transition on triangular lattices of sizes up to $N_c=16$. 
While the critical values of $U$ for the Mott transition are somewhat lattice-size dependent, the supercritical crossovers are the same for all values of $N_c$ at comparable temperatures. 
We thus argue that $N_c=6$ is sufficient to prove the existence of the Frenkel line. 

We focus mainly on the time-dependent conductivity, obtained from the Fourier transform of the real part $\sigma_1(\omega)$ of the conductivity. The latter is obtained from the imaginary part of the current autocorrelation function 
\begin{align}
    \sigma_1(t)&= \mathcal{F}\left[\sigma_1(\omega, \mathbf{q=0}) \right]=\mathcal{F}\left[\left\{\frac{\chi_{jj}''(\omega, \mathbf{q=0})}{\omega} \right\}\right],
\end{align} 
where $\mathcal{F}$ denotes Fourier transformation. 
Appendix~\ref{sec:FT} shows that  $\sigma^R(t)$ =$2 \mathcal{F}[\sigma_1(\omega)]$ for $t>0$.
We compute the current autocorrelation function from the bubble term and the first vertex correction. The bubble  term is
\begin{align}
    \chi_{j_{b_1}j_{b_1}}^{(0)}(iq_m,\mathbf{q = 0}) = \frac{-T}{N} \sum_{i\omega_n}\sum_{\kbf}  {v^2_{b_1}} \nonumber \\
    \times \Gmatsu(i\omega_n,\kbf)\Gmatsu(i\omega_n+iq_m, \kbf).
\end{align}
The velocity $v_{b_1}$ is taken in the direction of the reciprocal lattice vector $\hat{\mathbf{b}}_1 = \frac{\sqrt{3}}{2} \hat{x} + \frac{1}{2}\hat{y}$, namely
\begin{align}
    v_{b_1}=\hat{\mathbf{b}}_1\cdot\nabla_\mathbf{k} \epsilon_\mathbf{k} &= \sqrt{3}\left[t_1\sin(k_x)+ t_2\sin\left(\frac{k_x}{2}+\frac{\sqrt{3}k_y}{2}\right)\right].
\end{align}

In DCA, $\Gmatsu(i\omega_n,\kbf) = \sum_{i=0}^{N_c} \Gmatsu(i\omega_n, \Kbf_i)\phi_{\Kbf_i}(\kbf)$, where $\phi_{\Kbf_i}(\kbf) = 1$ if $\kbf$ is in the patch $\Kbf_i$, and 0 otherwise.
The first vertex correction, derived in Appendix~\ref{sec:vertex}, is 
\begin{align}
    &\chi_{j_{b_1}j_{b_1}}^{(1)}(iq_m,\mathbf{q = 0}) = \frac{-T}{N_c} \sum_{i\omega_n}\sum_{\kbf_{\text{border},ab}} v_{b_1} \\
    & \quad \times \Gmatsu_{ab}(i\omega_n,\kbf)\hat{\mathbf{b}}_1\cdot\mathbf{\Gamma}_{ab}(i\omega_n,i\omega_n+iq_m)\Gmatsu_{ab}(i\omega_n+iq_m, \kbf), \nonumber
\end{align}
where $a$ and $b$ label patches $0$ to $5$ and the sum on the wavevectors is only along the borders between pairs of patches $a$ and $b$ that are near-neighbor in Fig.~\ref{fig:clusterBZ}, and where we defined the average Matsubara Green's function on the borders by
\begin{align}
    \Gmatsu_{ab}(i\omega_n) &= \frac{\Gmatsu(i\omega_n, \Kbf_a)+\Gmatsu(i\omega_n, \Kbf_b)}{2},
\end{align}
and the three point vertex operator $\mathbf{\Gamma}_{ab}(i\omega_n,iq_m)$ by
\begin{align}
    \mathbf{\Gamma}_{ab}(i\omega_n,iq_m) &= \hat{\mathbf{n}}_{ab}\frac{\Gamma^{\text{vertex}}_{ab}(i\omega_n)+\Gamma^{\text{vertex}}_{ab}(i\omega_n+iq_m)}{2} \\
    \Gamma^{\text{vertex}}_{ab}(i\omega_n)&= \Sigma(i\omega_n,\Kbf_a)-\Sigma(i\omega_n,\Kbf_b),
\end{align}
where $\hat{\mathbf{n}}_{ab}$ is the unit vector perpendicular to the border between the patches $a$ and $b$.
This is inspired by, but differs from Ref.~\cite{lin_optical_2009}.

The Green's function, self-energy and current autocorrelation function are computed in Matsubara frequencies.
The analytic continuation and the computation of the spectral function of the current autocorrelation function $\sigma_1(\omega) = {\chi_{j j}''(\omega)}/{\omega}$  are obtained from the maximum entropy method~\cite{Jarrell:1996,bergeron_algorithms_2016,OmegaMaxEnt}.
Pad\'e approximants provide, in Appendix~\ref{sec:pade}, a check on the accuracy of the analytic continuation.
Further checks on the influence of the division in patches are provided in Appendix~\ref{sec:Glatt} while Appendix~\ref{sec:bubble} demonstrates that the Frenkel line is not drastically modified even if we do not include vertex corrections. The location of the critical point and critical slowing down are discussed in Appendix~\ref{sec:critical}.\\

\begin{figure}
    \centering
    \includegraphics[width=0.95\linewidth]{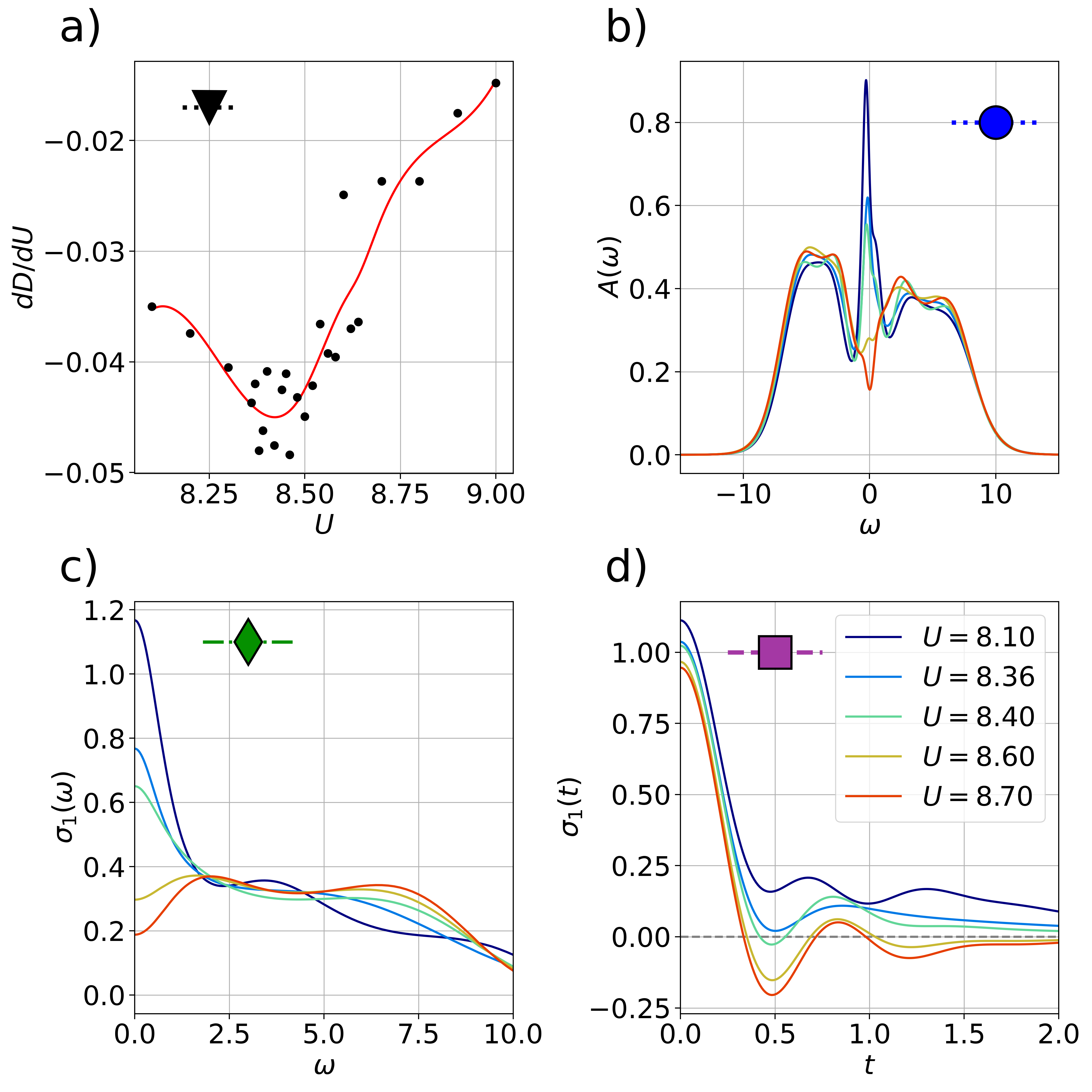}
    \caption{Definition of the Frenkel line and other crossover lines presented in Fig.~\ref{fig:phasediag}.
    All data in these figures are at $T=1/6$. 
    a) Derivative of the double occupancy with respect to the interaction $U$ as a function of $U$. %
    %
    The line of inflection of the double occupancy $U_D$ is given by the extrema of $dD/dU$. 
    b)~Density of states as a function of $\omega$ for the different values of $U$ found in the legend of Fig.~\ref{fig:differentlines}d). 
    A dip at $\omega=0$ appears at $U_{inc}$. It should be noted that under a sign flip of $t_2$, the density of states must be modified as follows $A(\omega)\to A(-\omega)$~\cite{downey_leffet_2022}.
    c) Optical conductivity as a function of $\omega$ for different values of $U$. 
    The Drude peak disappears at $U_{\sigma(\omega)}$. 
    d) Conductivity as a function of time $t$ for different  $U$. 
    That conductivity, $\sigma_1(t)$ can become negative at $U_{\sigma(t)}$. 
    This defines the Frenkel line. 
    The black triangle, blue circle, purple diamond, and red square represent the markers used to display the corresponding lines in  Fig~\ref{fig:phasediag}.}
    \label{fig:differentlines}
\end{figure}

\section{Results}
The data in Fig.~\ref{fig:differentlines} for various values of $U$ at fixed $T=1/6$ are used to define four crossover lines between the pseudogap and the correlated Fermi liquid: two that are known~\cite{sordi_pseudogap_2012,downey_mott_2023, downey_doping_2024}, and a third and fourth line based on the conductivity. These lines are drawn in the phase diagram in Fig.~\ref{fig:phasediag}.

The first and second crossover lines, respectively $U_D$ and $U_{inc}$ in Fig.~\ref{fig:phasediag}, have been published before~\cite{downey_mott_2023}.
Fig.~\ref{fig:differentlines}a) illustrates how the line of inflection in double occupancy, $U_D$, is obtained from the minima of $dD/dU$. This is a thermodynamic line. 
Fig.~\ref{fig:differentlines}b) shows how the appearance of a dip in $A(\omega)$ marks the estimate of the dynamic crossover line $U_{inc}$. The dip is found between $U=8.60$ and $U=8.70$.

The third crossover line is obtained from the loss of the Drude peak centered at $\omega=0$ in the frequency-dependent conductivity $\sigma_1(\omega)$~\cite{orenstein_frequency-_1990,homes_optical_1993,timusk_pseudogap_1999}. 
In Fig.~\ref{fig:differentlines}c) the dip in $\sigma_1(\omega)$ near $\omega=0$ marks our estimate of this crossover line. The dip is found between $U=8.45$ and $U=8.60$. We label this line $U_{\sigma(\omega)}$

The last crossover line is obtained from the Fourier transform of the optical conductivity $\sigma_1(t)$. In Fig~\ref{fig:differentlines}d), the appearance of negative values of $\sigma_1(t)$  marks our estimate of this crossover line. These negative values occur between $U=8.40$ and $U=8.45$. We label this line $U_{\sigma(t)}$. This is our definition of the electronic Frenkel line.

These four crossover lines are plotted in Fig~\ref{fig:phasediag} on the basic phase diagram previously published in Ref.~\citenum{downey_mott_2023}. This is the $T-U$ phase diagram for the half-filled triangular lattice Hubbard model.
The four crossover lines $U_{\sigma(t)}$, $U_{\sigma(\omega)}$, $U_D$, and $U_{inc}$ in the supercritical region separate the pseudogap and metallic regimes. They are illustrated with the dotted and dashed lines. The critical point~(Appendix~\ref{sec:critical}) very near $T_c=0.112$, $U_c = 8.258$ is labelled by the red star.

At temperatures below the critical point, the black solid line represents a first-order phase transition: A discontinuity is found in the double occupancy, the density of states, $\sigma_1(\omega)$ and $\sigma_1(t)$ at the same values of $U$.
The hatched region between $U_{c1}$ and $U_{c2}$ is a region of phase coexistence, while the dashed line $U_{gap}$~\cite{downey_mott_2023} represents the crossover between the pseudogap and Mott insulator. All crossover lines between the pseudogap and correlated Fermi liquid emerge from the critical point.

As the temperature increases, the extremum in $\dd D/\dd U$ that defines $U_D$ becomes less pronounced and is eventually obscured by numerical noise. If we assume that the $U_D$ line disappears when the relative difference between $\dd D/\dd U$ at the extremum and at $U = U_{D} \pm 0.1$ becomes less than $0.5\%$, the $U_D$ line disappears at $T \geq 0.29$.\\

\section{Discussion}
The electronic Frenkel line, which we defined here, separates the pseudogap and correlated Fermi liquid in the vicinity of $U_{D}$, the inflection of double occupancy as a function of $U$. It is also close to the two dynamic lines $U_{inc}$ and $U_{\sigma(\omega)}$. This means that the appearance of negative $\sigma_1(t)$ is a legitimate way to define the crossover from the pseudogap to the correlated Fermi-liquid regimes. 

The electronic Frenkel line emerges from the Mott critical point. This differs from the Frenkel line found in classical fluids, where it emerges from a point at a temperature below $T_c$, and lies in the liquid phase for $T<T_c$ \cite{pruteanu_frenkel_2021,bryk_behavior_2017,brazhkin_two_2012,yang_frenkel_2015}. This indicates that, in classical fluids, structural and dynamic lines are different. In our case, we study a transition between the pseudogap and correlated Fermi liquid, which is dynamical rather than structural. That the Frenkel line starts at the Mott critical point is expected because this point reflects a dynamical, rather than structural, form of criticality.

Another difference between the electronic Frenkel line and that found in classical fluids is that, as explained in the introduction, the crossover criterion differs. In classical fluids, the criterion for the dynamic line is the temperature and pressure where the local minima and oscillations disappear in the velocity autocorrelation function \cite{brazhkin_two_2012}. This criterion for the Frenkel line in classical fluids comes from the monotonic decay of the velocity autocorrelation function in the gas phase. In the triangular-lattice Hubbard model, we do not find any temperature or interaction $U$ for which the autocorrelation function does not present oscillations or local minima. 

Furthermore, the interaction responsible for the change of dynamics across the Frenkel line differs between electronic and classical fluids.
In classical fluids, the crossing of the Frenkel line is associated with the disappearance of transverse {\it acoustic} phonon modes \cite{trachenko_theory_2023,bolmatov_frenkel_2015}. Here, crossing the Frenkel line by decreasing $U$ corresponds to the disappearance of back-scattering. 

Experimentally, in cuprates, peaks associated with {\it optical} phonons are found in  optical conductivity $\sigma_1(\omega)$ at energies between $60~\text{cm}^{-1}$ and $700~\text{cm}^{-1}$ \cite{mirzaei_spectroscopic_2013,laforge_sum_2008,basov_sum_1999}.
Therefore, to experimentally observe the proposed Frenkel line, it is necessary to suppress or filter optical-phonon contributions to the signal.

The advantage of the electronic Frenkel line over the dynamic lines $U_{\sigma(\omega)}$ and $U_{inc}$ is that it is based on a sharp criterion: negative $\sigma_1(t)$. Contrary to the thermodynamic line $U_D$, it extends to very high temperatures. It marks a real qualitative change in the dynamics of the electrons, namely the appearance of back-scattering. In contrast with all other lines, this negative $\sigma_1(t)$ also reflects high-frequency behavior. \\ 

\section{Conclusion}
We report the existence of an electronic Frenkel line, analogous to the Frenkel line in classical fluids, that can be used to separate the pseudogap and metallic regimes of the supercritical state. The Frenkel line in the electronic problem is given by a sharp criterion, namely, negative $\sigma_1(t)$. It is found in the vicinity of other lines used to define the pseudogap state in the phase diagram. It emerges from the critical point and extends to high temperatures. Comparison with experiments calls for subtraction of phonons and measurements over a wide frequency range. We did not find data satisfying both requirements. 
However, measurements of the Frenkel line in cold-atom quantum simulators should be possible. 

While we focused on the half-filled triangular-lattice Hubbard model, a critical point and pseudogap regime are also found at finite doping \cite{downey_doping_2024,fournier_two_2024,gull_superconductivity_2013}, as well as for the square-lattice Hubbard model \cite{semon_continuous-time_2014,kyung_mott_2006,braganca_correlation-driven_2018}. We speculate that the electronic Frenkel line is a general concept that will also be useful for these cases and other electronic phase transitions.\\

\acknowledgments
Useful discussions with A. Georges, R. Lobo, K. Trachenko and D. van der Marel are acknowledged. This work has
been supported by the Natural Sciences and Engineering Research Council of Canada (NSERC) under grant
RGPIN-2024-05206 and by the Canada First Research Excellence Fund. The computational resources were provided by Calcul-Québec and the Digital Research Alliance of Canada.
The Flatiron Institute is a division of the Simons Foundation.

\newpage
\appendix

\section{Fourier transform of the retarded conductivity}
\label{sec:FT} 
The retarded response in frequency has both a real ($\sigma_1(\omega)$) and an imaginary ($\sigma_2(\omega)$) part so that its Fourier transform is given by 
\begin{align}
    \sigma^R(t) &=  \left[\int_{-\infty}^\infty \left(\sigma_1(\omega) + i\sigma_2(\omega) \right) \Theta(t) e^{-i \omega t} d\omega \right].
\end{align}
with $\Theta(t)$ the Heaviside function. The imaginary part of the optical conductivity is obtained through the Kramers-Kronig relation.

In mathematics, the Kramers-Kronig relation is a Hilbert transformation. 
The following property of Hilbert transformations can be used to find $\sigma^R(t)$ using only $\sigma_1(\omega)$. 
\begin{align}
    \mathcal{F}\left[H(u) \right](t) = - i\cdot\text{sgn}(t)\cdot \mathcal{F}[u](t),
\end{align}
where $\mathcal{F}$ stands for Fourier transform, $H$ for Hilbert transformation, and $u$ is a real function of frequency. This formula can also be proven using Laplace transforms. Inserting this relation in the formula for $\sigma^R(t)$, we find
\begin{align}
    \sigma^R(t) 
    &=  \left[\mathcal{F}[\sigma_1(\omega)\right] + i \mathcal{F}\left[H\left[ \sigma_1(\omega)] \right]\right]\\
    &=  \mathcal{F}\left[\sigma_1 \right](t) \left\{1+ \text{sgn}(t) \right\}.
\end{align}
Thus, $\sigma^R(t)$ equals $2[\mathcal{F}[\sigma_1]]\Theta(t)$ and we do not need to compute $\sigma_2(\omega)$ nor $\sigma_2(t)$. Also, as required, $\sigma^R(t)$ is real.

\section{Optical conductivity and vertex correction in DCA}\label{sec:vertex}
We start from the equation for the current autocorrelation function
\begin{align}
    & \chi_{j_{b_1} j_{b_1}}(iq_m, \mathbf{q=0}) = \\ 
    & \nonumber \quad \text{Tr} \left[\frac{\partial \epsilon_\kbf}{dk_{b_1}}\Gmatsu(i\omega_n,\kbf)\hat{\mathbf{b}}_1 \cdot \mathbf{\Gamma}(i\omega_n,iq_n,\kbf)\Gmatsu(i\omega_n+iq_m,\kbf) \right].
\end{align}
%
The three-point current vertex $\mathbf{\Gamma}$ can be obtained from the Ward identity~\cite{bergeron_optical_2011,lin_optical_2009}
\begin{align}
    iq_m\Gamma_\rho - \mathbf{q}\cdot \mathbf{\Gamma} &= \Gmatsu^{-1}(i\omega_n+iq_m,\kbf+\qbf)-\Gmatsu^{-1}(i\omega_n,\kbf), \label{eq:WardId}
\end{align}
%
where $\Gamma_\rho$ is the three-point charge vertex.
In the $\qbf \rightarrow 0$ limit, the inverse Matsubara Green's function can be approximated with a Taylor expansion.
\begin{multline}
    \label{eq:small_q_Ginv}
    \lim_{\qbf \rightarrow 0} \Gmatsu^{-1}(i\omega_n+iq_m, \kbf+\qbf) \approx \Gmatsu^{-1}(i\omega_n+iq_m,\kbf)\\+ \frac{\delta \Gmatsu^{-1}(i\omega_n+iq_m,\kbf)}{\delta \kbf} \cdot \qbf
\end{multline}
By inserting the expressions for $\Gmatsu^{-1}(\kbf+\qbf,i\omega_n+iq_m)$ and $\Gmatsu^{-1}(\kbf,i\omega_n)$  in the Ward identity Eq.~(\ref{eq:WardId}) and employing the above equation and the Dyson equation, we find 
\begin{multline}
    iq_m\Gamma_\rho - \mathbf{q}\cdot \mathbf{\Gamma} = iq_m - \left[ \Sigma( i\omega_n+iq_m,\kbf+\qbf) - \Sigma( i\omega_n, \kbf)\right] \\
    + \frac{\delta \Gmatsu^{-1}(i\omega_n+iq_m,\kbf)}{\delta \kbf} \cdot \qbf.
\end{multline}
In the limit $\qbf \rightarrow 0$, the scalar part of the three-point function, related to density, is given by
\begin{multline}
    iq_m \Gamma_\rho(i\omega_n, iq_m, \kbf) =\\ iq_m - \left[ \Sigma( i\omega_n+iq_m,\kbf) - \Sigma(i\omega_n,\kbf)\right].
\end{multline}
The current vertex is thus given by
\begin{align}
    \qbf \cdot \mathbf{\Gamma} &= -\frac{\delta \Gmatsu^{-1}(i\omega_n+iq_m,\kbf)}{\delta \kbf} \cdot \qbf\\
    &= \qbf \cdot \left[\frac{\delta \epsilon(\kbf)}{\delta \kbf} + \frac{\delta \Sigma(i\omega_n+iq_m,\kbf)}{\delta \kbf}\right].
\end{align}
The first term, $\mathbf{q}\cdot \frac{\delta \epsilon(\kbf)}{\delta \kbf}$, leads to the bubble contribution, while the second term $\mathbf{\Gamma}^{\text{vertex}} = \delta \Sigma(i\omega_n+iq_m,\kbf)/{\delta \kbf}$ is a vertex correction for $\chi_{j j}(iq_m)$.
DCA is a coarse-grained method in $\kbf$-space, and the self-energy is constant within the patches of the Brillouin zone. The derivative of the self-energy is nonzero only at the border of every patch~\cite{lin_optical_2009}. The three-point vertex function is thus
\begin{align}
    \mathbf{\Gamma}^{\text{vertex}}_{ab}(i\omega_n,iq_m,\kbf) =  & \left[ \Sigma^b(i\omega_n+iq_m)-\Sigma^a(i\omega_m+iq_m)\right] \nonumber \\
    & \quad \times \delta([\kbf-\kbf_{ab}]\cdot \mathbf{\hat{n}}_{ab})\mathbf{\hat{n}}_{ab} ,
\end{align}
where $\mathbf{\hat{n}}_{ab}$ is the normal vector to the border between the patches labelled $a$ and $b$, and $\kbf_{ab}$ is the vector defining the border between the two patches. To obtain a real current autocorrelation function, we need to symmetrize the three-point vertex in the equation for  $\chi_{jj}
$.
\begin{align}
    \mathbf{\Gamma}_{ab}(\kbf,i\omega_n,iq_m) &= \frac{\mathbf{\Gamma}^{\text{vertex}}_{ab}(\kbf,i\omega_n)+ \mathbf{\Gamma}^{\text{vertex}}_{ab}(\kbf,i\omega_n+iq_m)}{2}. 
\end{align}
The same result follows without need for symmetrization in frequency if, in  Eq.~(\ref{eq:WardId}), we write the right-hand side as $\Gmatsu^{-1}(i\omega_n+iq_m,\kbf+\frac{\qbf}{2})-\Gmatsu^{-1}(i\omega_n,\kbf-\frac{\qbf}{2})$.

This approach assumes that the contributions ${\Gamma}_\rho$ and $\mathbf{\Gamma}$ are both analytic in $\qbf$ and $i\omega_n$. It leads to the so-called Maki-Thomson part of the vertex corrections and misses the Aslamasov-Larkin part~\cite{bergeron_optical_2011}. The effect of vertex corrections is discussed further in Appendix~\ref{sec:bubble}.


\section{Comparison with Padé approximant}\label{sec:pade}

\begin{figure}
    \centering
    \includegraphics[width=0.95\linewidth]{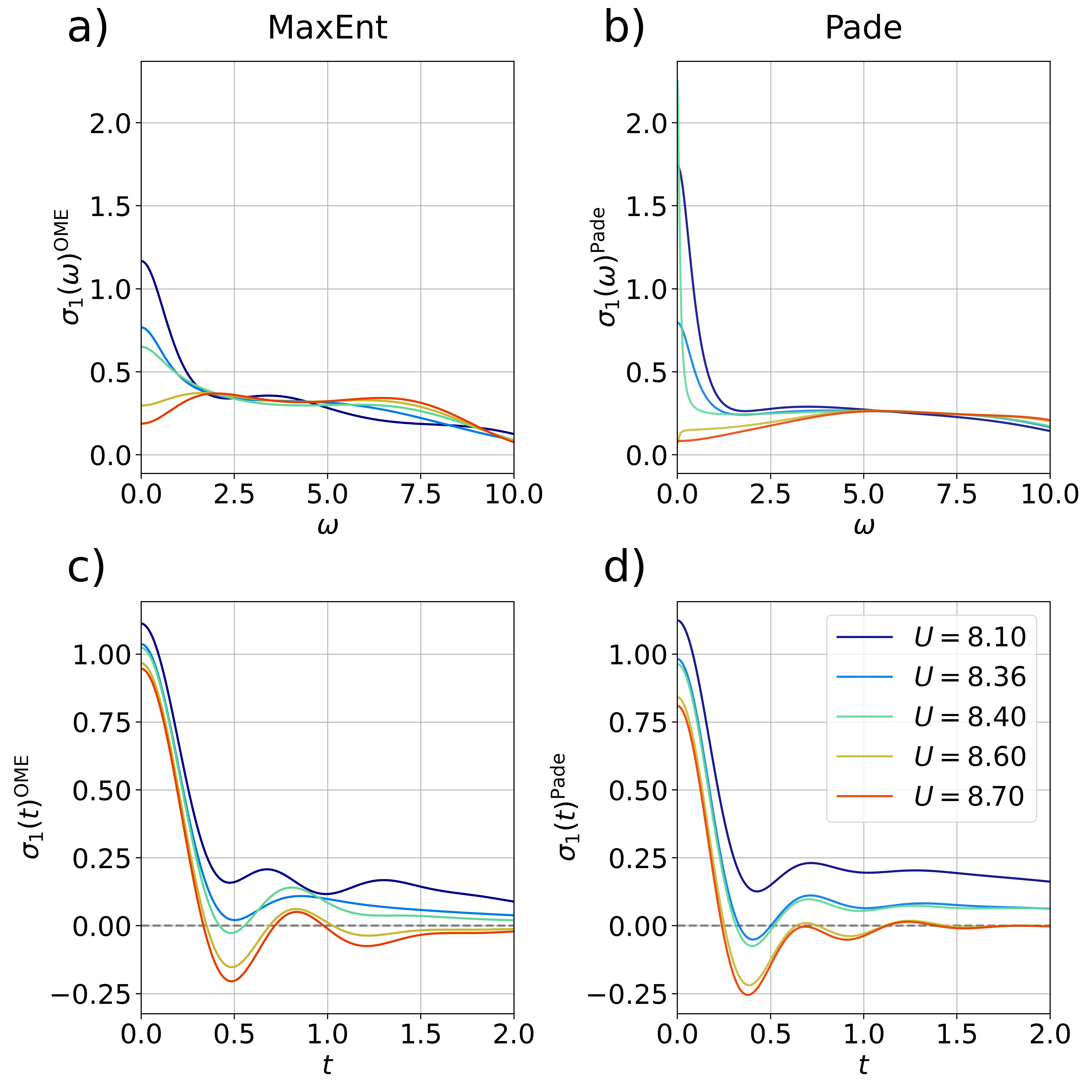}
    \caption{a) Frequency-dependent optical conductivity obtained from analytical continuation with the maximum entropy method OmegaMaxEnt~\cite{OmegaMaxEnt}; and b) with Padé approximants. 
    c) Time-dependent optical conductivity obtained from Fourier transformation of $\sigma_1(\omega)$ in a); and d) Fourier transformation of $\sigma_1(\omega)$ in b). All these curves are for $T=1/6$.}
    \label{fig:OMEvsPade}
\end{figure}

Extracting real-frequency behaviour of the optical conductivity from Matsubara frequency data requires analytic continuation. This is a numerically ill-posed inverse problem. In the main text, the analytic continuation is performed with the Bayesian so-called ``maximum entropy'' method~\cite{Jarrell:1996} using the publicly available software OmegaMaxEnt~\cite{bergeron_algorithms_2016,OmegaMaxEnt}. In this Appendix, we assess the robustness of our results by comparing them with results from Padé approximants. That method can be used only when statistical errors from the continuous-time quantum Monte Carlo are very small.

Results for the frequency- and time-dependent conductivities, obtained with OmegaMaxEnt and with Padé approximants, are displayed in Fig.~\ref{fig:OMEvsPade}. For the optical conductivity, we find that both methods give very similar results for the Drude response near $\omega\rightarrow 0$. Moreover, both methods show a maximum in the mid-frequency range of $\sigma_1(\omega)$, although there is more structure with the MaxEnt method.

Moreover, we observed an overestimation of the Drude peak for $U=8.40$ near the crossover line $U_{\sigma_1(\omega)}$. This is because Padé approximants are sensitive to numerical errors at low Matsubara frequencies. Such sensitivities are further amplified near the crossover regime, where the optical response changes rapidly with interaction strength.

However, most importantly, we find that despite the quantitative differences between $\sigma_1(\omega)$ and $\sigma_1(t)$ obtained with the two methods, both show a suppression of the Drude peak and the appearance of negative time-dependent conductivity at large $U$. Moreover, these crossovers are found for similar values of interaction for both methods. This indicates that the Frenkel line found in this work appears to be a genuine feature of the optical response, rather than an artifact of the analytic continuation.



\section{$\Gmatsu_{\text{DCA}}$ vs $\Gmatsu_{\text{lat}}$}\label{sec:Glatt} 
\begin{figure}
    \centering
    \includegraphics[width=0.95\linewidth]{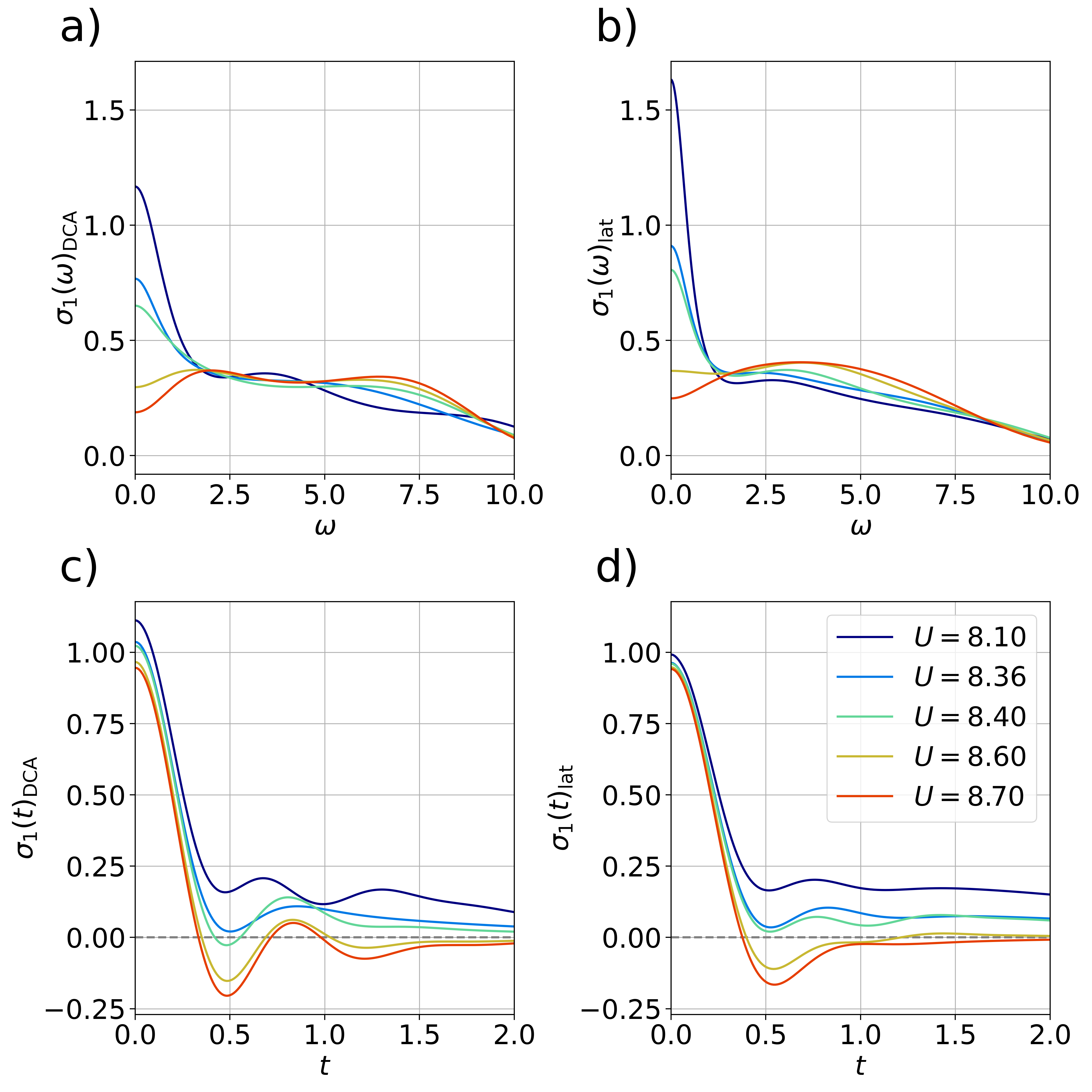}
    \caption{a) Frequency-dependent optical conductivity obtained with the DCA Green's function, and b) with the lattice Green's function. c), d) Time-dependent optical conductivity obtained from Fourier transformation of $\sigma_1(\omega)$ in a); and d) Fourier transformation of $\sigma_1(\omega)$ in b). All these curves are for $T=1/6$.}
    \label{fig:Glat}
\end{figure}

In this letter, we use $\Gmatsu_{\text{DCA}}$ to compute the current autocorrelation function. $\Gmatsu_{\text{DCA}}$ is a coarse-grained Green's function constant within each patch of the Brillouin zone. However, in DCA, it is also possible to compute the lattice Green's function
\begin{align}
    \Gmatsu_{\text{lat}}(\Kbf,\Tilde{\mathbf{k}},i\omega_n) &= \left(i\omega_n-\epsilon_{\Kbf + \Tilde{\mathbf{k}}}+\mu-\Sigma(\Kbf)\right)^{-1}
\end{align}
where $\mathbf{\Tilde{k}}$ is the wavector inside the different patches. Using this definition of the Green's function changes the sum over the wave vectors in the computation of $\chi_{jj}$. Fig.~\ref{fig:Glat} shows the results for $\sigma_1(\omega)$ and $\sigma_1(t)$ obtained with the DCA Green's function and the lattice Green's function. Even though the Drude peak is sensitive to the choice, we find that the different Green's functions give overall similar results for the conductivity, and, most importantly, that the location of the Frenkel line does not depend significantly on the Green's function used. 

\section{Bubble term vs vertex corrections}\label{sec:bubble}

\begin{figure}
    \centering
    \includegraphics[width=0.95\linewidth]{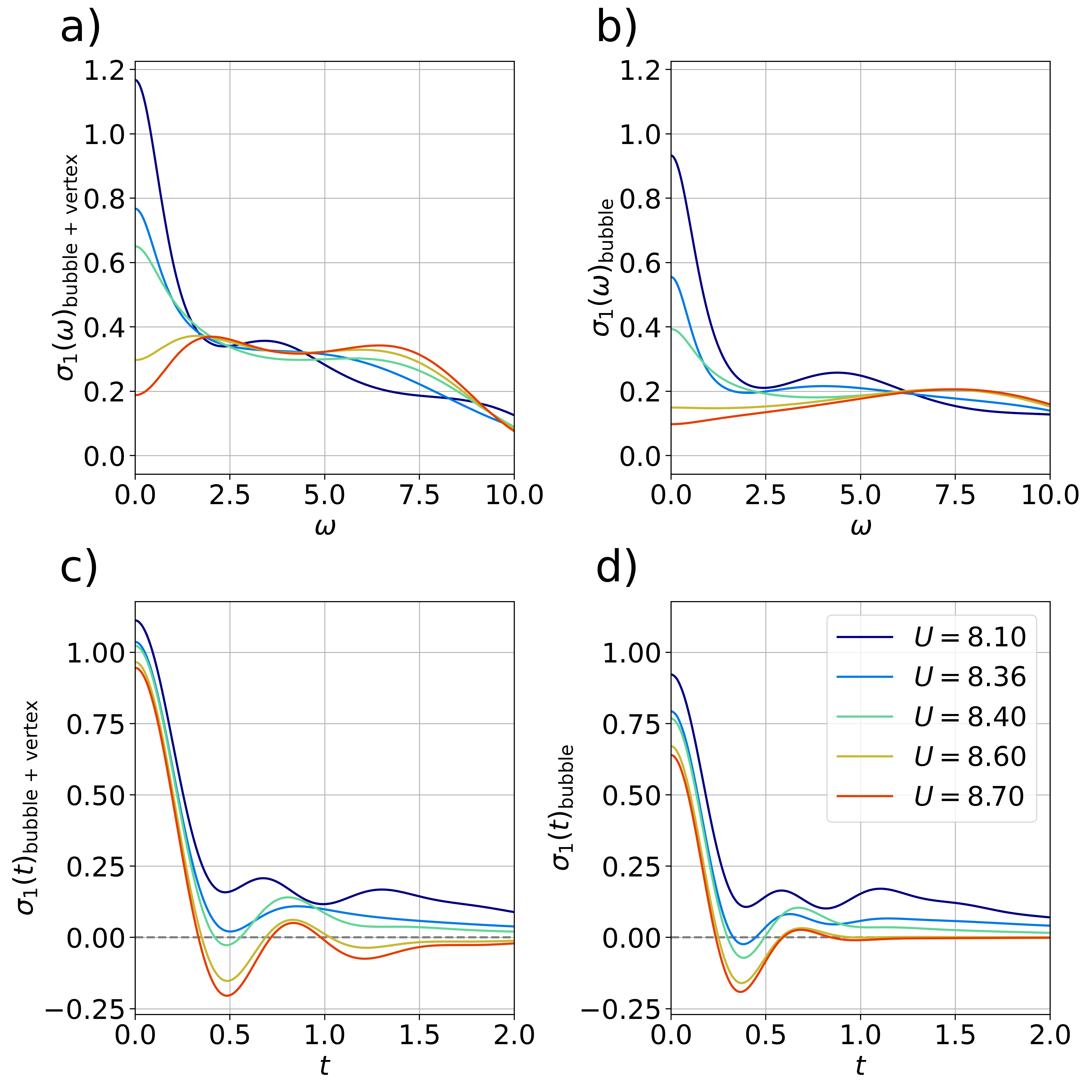}
    \caption{a) Frequency-dependent optical conductivity included the bubble term and the first vertex correction from the Ward identity, and b) only the bubble term. 
    c) Time-dependent optical conductivity obtained from Fourier transformation of $\sigma_1(\omega)$ in a); and d) Fourier transformation of $\sigma_1(\omega)$ in b). All these curves are for $T=1/6$.}
    \label{fig:optcond_bubble}
\end{figure}

\begin{figure}
    \centering
    \includegraphics[width=0.9\linewidth]{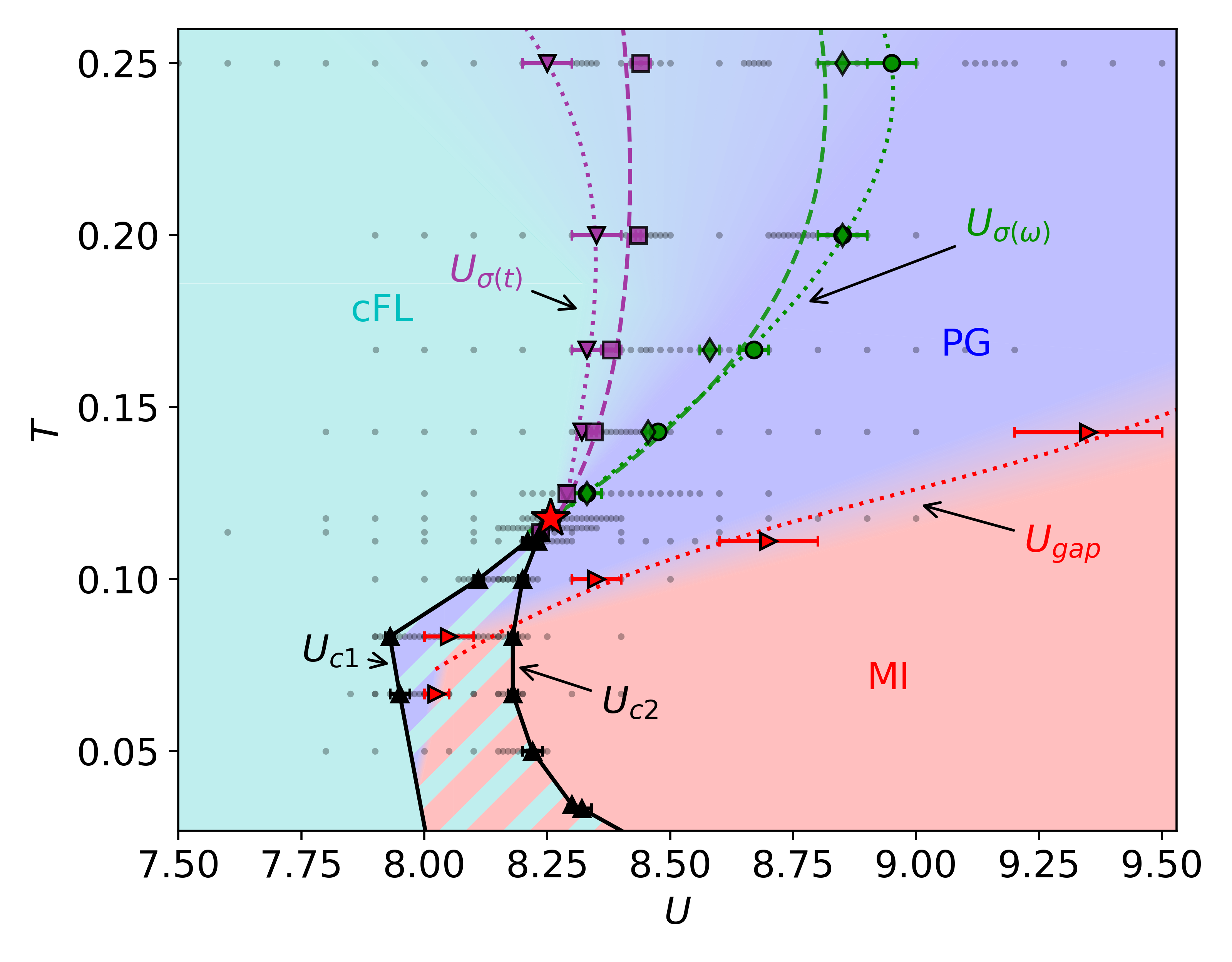}
    \caption{$T-U$ phase diagram of the triangular-lattice Hubbard model at half-filling, where $U_{\sigma(t)}$ and $U_{\sigma_1(\omega)}$ dotted lines with pointers are obtained taking only the bubble term in the optical conductivity, and the dashed lines are obtained by taking the bubble term and the first vertex correction for the optical conductivity. 
    The other lines, and their definitions, are unchanged from the original Fig.~\ref{fig:phasediag}. The maximum temperature of the phase diagram is $T=0.25$, different from $T=0.34$ in Fig.~\ref{fig:phasediag}.}
    \label{fig:phasediagbubble}
\end{figure}

Throughout, the optical conductivity used to find the lines $U_{\sigma(\omega)}$ and the Frenkel line $U_{\sigma(t)}$ includes the bubble term and the first vertex corrections from the Ward identity, as discussed in Appendix~\ref{sec:vertex}. In this Appendix, we compare our results for $\sigma_1(\omega)$, $\sigma_1(t)$, and the phase diagram in Fig.~\ref{fig:phasediag} from the main text with those obtained when the optical conductivity is evaluated at the bubble level only. The purpose of this analysis is (i) to verify that the concept of the Frenkel line remains defined without vertex corrections, (ii) to assess the shift of the lines $U_{\sigma(\omega)}$ and the Frenkel line $U_{\sigma(t)}$ when vertex corrections are neglected and (iii) to study the sensitivity of the Frenkel line to vertex corrections.

First, we present the results for the time- and frequency-dependent conductivity computed with the bubble and the first-order vertex correction at $T=0.1667$. These are displayed in Fig.~\ref{fig:optcond_bubble}. We observe that the addition of the vertex correction enhances the optical conductivity at $\omega \sim 2.5t$, especially in the weak and intermediate interaction regimes. However, the vertex correction does not significantly modify the quantitative behaviour of the $\omega \rightarrow 0$ optical conductivity. For the time-dependent conductivity, we observe that the addition of vertex corrections suppresses the oscillations in the signal. However, it does not drastically alter the shape of the conductivity.

Having established the quantitative impact of the vertex corrections on $\sigma_1(\omega)$ and $\sigma_1(t)$, we turn our attention to their impact on the crossover lines $U_{\sigma(\omega)}$ and $U_{\sigma(t)}$. Fig.~\ref{fig:phasediagbubble} presents a modified phase diagram. The lines $U_{\sigma(\omega)}$ and $U_{\sigma(t)}$ are defined using the optical conductivity at the bubble level. This new phase diagram is restricted to $T<0.26$, compared to $T<0.35$ in the main text. This limitation arises from analytic continuation, which becomes unreliable here for large $U$ at temperatures higher than $T=0.25$, preventing us from finding the Frenkel line.\\

Looking at Fig.~\ref{fig:phasediagbubble}, we can address the three points raised earlier in the Appendix. First, we find that, even when the optical conductivity is computed at the bubble level, the lines $U_{\sigma(\omega)}$ and $U_{\sigma(t)}$ appear in the supercritical region of the phase diagram, originating from the Mott critical point $(U_c, T_c)$ (Appendix~\ref{sec:critical}). We had found the same overall behavior with vertex corrections. 
Moreover, at temperatures lower than $T_c$, there is a coexistence of two solutions for both $\sigma(\omega)$ and $\sigma(t)$ between $U_{c_1}$ and $U_{c_2}$. Thus, even at the bubble level, our criteria for $U_{\sigma(\omega)}$ and $U_{\sigma(t)}$ are reliable to identify the first-order phase transition between the correlated metal and the pseudogap, as well as the crossover regime in the supercritical region. Also, we see that the positions of the Frenkel line $U_{\sigma(t)}$ and of $U_{\sigma(\omega)}$ are still in the vicinity of $U_D$ and $U_{inc}$. This indicates that the bubble term controls the overall behaviour of the optical conductivity. In this work, the only vertex corrections included are obtained from the Ward identity. The corrections of the type Aslamazov-Larkin~\cite{bergeron_optical_2011} are not included. However, since the Frenkel line is already well captured at the bubble level and only weakly affected by the Maki-Thomson-like vertex corrections from the Ward identity, it is reasonable to expect that additional vertex corrections would not qualitatively modify the location of the Frenkel line. Nevertheless, this remains an open problem.


\section{Critical point}\label{sec:critical} 
\begin{figure}
    \centering
    \includegraphics[width=0.95\linewidth]{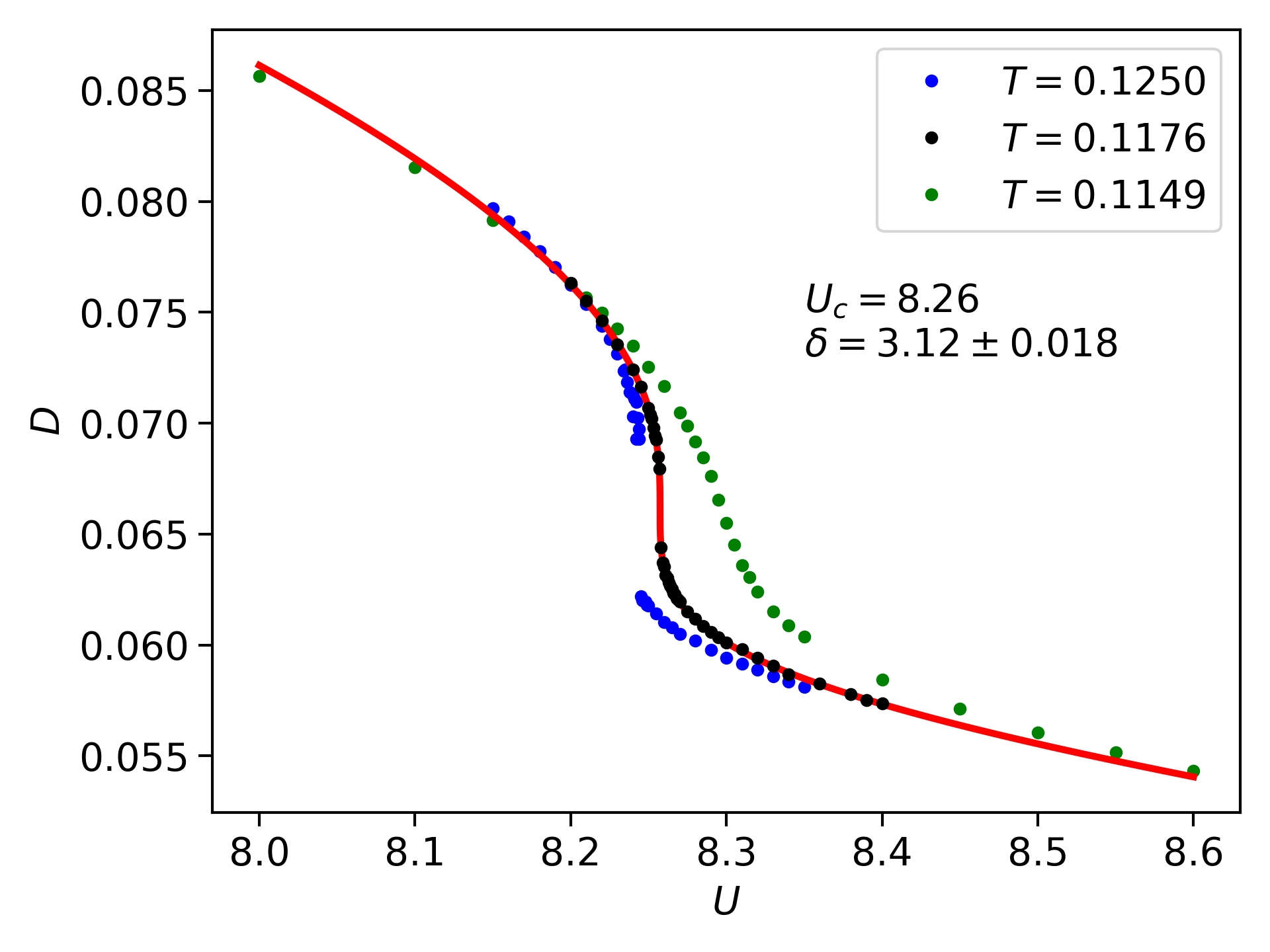}
    \caption{Double occupancy as a function of $U$ at $T=0.1250$, $0.1176$ and $0.1149$ (or inverse temperature $\beta = 8.0$, $8.5$ and $8.7$). Eq.~ (\ref{eq:Dc}) is used to fit $D$ in order to find the critical occupancy $D_c$, interaction $U_c$ and critical exponent $\delta$ at the critical temperature $T_c=0.1176$. The values of $U_c$ and $\delta$ obtained with the fit are included in the figure.}
    \label{fig:doccUCriticalPoint}
\end{figure}

The Mott critical point can be found from the double occupancy as a function of interaction $U$ for different temperatures. At temperature $T<T_c$, there is a first-order transition for the double occupancy at $U_{c1}$ and $U_{c2}$, whereas at $T>T_c$, the double occupancy is continuous. The double occupancy as a function of $U$ for the inverse temperatures $\beta=8.0$, $\beta=8.5$ and $\beta=8.7$ is shown in Fig.~\ref{fig:doccUCriticalPoint}. The temperature $T = 1/8.5 \ approx 0.1176$ is the largest of these three temperatures for which we find a discontinuous double occupancy. We found a continuous double occupancy at $T=1/8=0.1250$. 

Following Ref.~\cite{semon_importance_2012}, we obtain the value of the critical interaction $U_c$  by fitting the data for the double occupancy  $U$ at $\beta_c = 8.5$ with the function
\begin{align}
    D-D_c &= c_1 \text{sgn}(\delta U)|\delta U|^{1/\delta} + c_2 |\delta U|^{2/\delta} + c_3 \delta U \label{eq:Dc},
\end{align}
where $\delta U  = U-U_c$, and $D_c$, $U_c$, $\delta$, $c_1$, $c_2$ and $c_3$ are fitting parameters .

\begin{figure}
    \centering
    \includegraphics[width=0.95\linewidth]{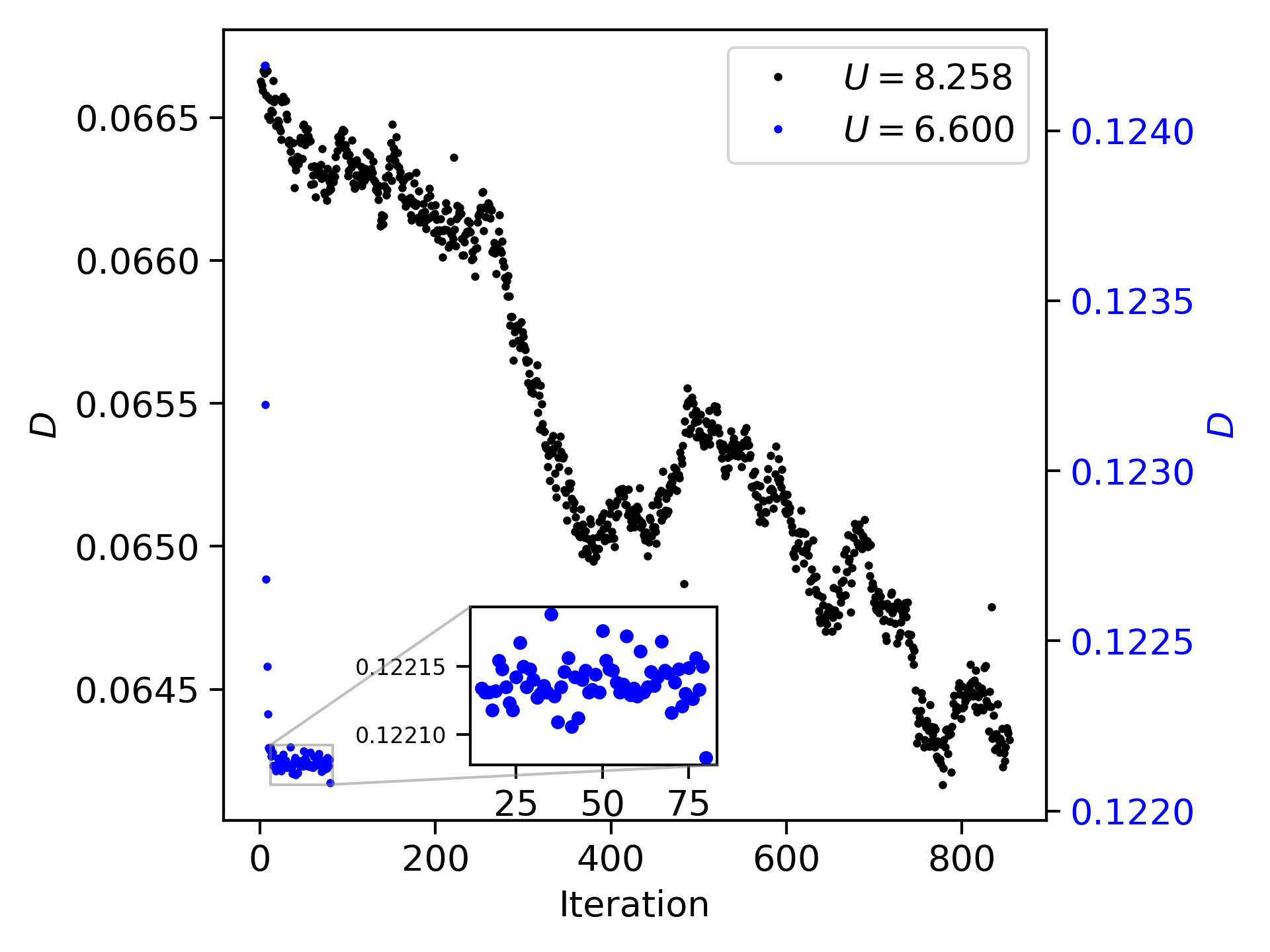}
    \caption{Double occupancy as a function of iteration for $U=6.600$ in blue and $U=8.258$ in black at the critical temperature $\beta=8.5$. The inset is a zoom of $D$ as a function of iterations starting at the tenth iteration.}
    \label{fig:convergence}
\end{figure}
Another method to verify the accuracy of the critical point is to study the convergence of each simulation. At the critical point, the correlation length diverges. This leads to critical slowing down, which causes convergence problems. At $\beta_c = 8.5$, and $U=8.257$, it takes more than 800 iterations for the double occupancy to converge. This is illustrated in Fig.~\ref{fig:convergence}, where we show double occupancy as a function of iteration for $U=8.258$ and $U=6.600$. We see that at $U=6.600$, far from the critical point, the DCA calculation converges in less than 10 iterations, and there are very small fluctuations in subsequent iterations. At $U=8.258$, the calculation is still not converged after 800 iterations. This supports the claim that $\beta_c = 8.5$, $U=8.257$ is very close to the critical point. As the small jump in the red solid line of the double occupancy fit in Fig.~\ref{fig:doccUCriticalPoint} suggests, our estimate for the critical temperature is probably slightly too low, leading to some small hysteresis in addition to critical slowing down.

\end{document}